\documentclass[journal]{IEEEtran}
\usepackage[T1]{fontenc}
\usepackage[utf8]{inputenc}
\usepackage{newtxtext,newtxmath}
\usepackage{amsmath}
\usepackage{graphicx}
\usepackage{booktabs,tabularx,array}
\usepackage{cite}
\usepackage{xurl}
\usepackage{listings}
\usepackage{xcolor}
\usepackage{microtype}
\usepackage[hidelinks]{hyperref}
\newcolumntype{Y}{>{\raggedright\arraybackslash}X}
\lstdefinestyle{ciyaml}{basicstyle=\ttfamily\scriptsize,breaklines=true,
  columns=fullflexible,keepspaces=true,frame=single,rulecolor=\color{black!25},
  showstringspaces=false,aboveskip=6pt,belowskip=6pt}
\hypersetup{
  pdftitle={Capability Internet: A Network Architecture for Autonomous Agents},
  pdfauthor={Zhuoming WEN; Hongwei LI; You WU},
  pdfsubject={An agent-oriented Internet architecture for capability exchange},
  pdfkeywords={Capability Internet, Internet architecture, autonomous agents, capability exchange, task coordination, federated resources}}

\title{Capability Internet:\\A Network Architecture for Autonomous Agents}
\author{Zhuoming~WEN, Hongwei~LI\textsuperscript{*}, and You~WU\textsuperscript{*}%
\thanks{Zhuoming WEN is with STAR.VISION Aerospace Group Limited, Hangzhou 310027, China (e-mail: \href{mailto:wen.zhuoming@star.vision}{wen.zhuoming@star.vision}).}%
\thanks{Hongwei LI and You WU are with Zhejiang University, Hangzhou 310027, China. \textsuperscript{*}Corresponding authors: Hongwei LI and You WU (e-mail: \href{mailto:lhw1992@zju.edu.cn}{lhw1992@zju.edu.cn}; \href{mailto:12524104@zju.edu.cn}{12524104@zju.edu.cn}).}}

\begin{document}
\maketitle

\begin{abstract}
Autonomous agents increasingly provide and consume network services. An agent coordinating a long-running task across organizations must track what has been done, which actions are authorized, and which provider can perform the next step. We propose Capability Internet (CI), an architecture in which agents exchange capabilities and coordinate their use through persistent tasks. Versioned descriptions specify behavior, and live instances identify the providers and resources that can supply it. Task records link each invocation to its selected instance, authority, and terms of use. When providers change, the task can reuse completed results that still meet its requirements. Five functions support capability discovery, task coordination, federated compute, intelligence reuse, and governance of capability use and settlement. Reviewed task results can supply versioned intelligence for subsequent tasks. We illustrate CI through four deployment profiles covering computing and intelligence services, manufacturing, vehicles and embodied robots, and space--ground operations, and propose an evaluation plan.
\end{abstract}


\section{Introduction}
\label{sec:intro}

Over the past three decades, the Internet has undergone three shifts in function. The early \emph{Information Internet} was designed to help people exchange information through email, web pages \cite{www1994}, search, video, and social media. In the second stage, the Internet shifted toward \emph{cloud computing}. As computing resources became concentrated in large data centers, the Internet began to offer computing capacity in addition to carrying information \cite{ref5,serverless2023}. In both stages, the network remained centered on people and the applications that served them.

The third shift concerns who participates in network services. Large numbers of autonomous software agents, robots, vehicles, and industrial devices are becoming new users of the Internet. They can provide services themselves and invoke services on behalf of individuals and organizations \cite{autogen2024,ioaSurvey2026}. These systems pursue their goals by obtaining resources and coordinating actions across providers.

What they exchange over the network is also changing. An agent combines operations such as model inference, knowledge retrieval, navigation, robot control, and payment to carry out a task. Each operation must be discoverable, authorized for use, and able to supply what later steps require.

A wildfire-response task combines satellite imaging, ground analysis, and the dispatch of response assets. Ground analysis needs a timely image in a format it can process. The response team needs the resulting plan and authority to act. The coordinator tracks those dependencies across organizations as each stage finishes.

Internet of Agents (IoA) supports discovery and dynamic teaming among heterogeneous third-party agents \cite{ioa2025}. Agent2Agent (A2A) specifies agent interactions and tasks, while the Model Context Protocol (MCP) exposes tools and resources \cite{ref8,ref9}. To coordinate work across providers, participants also need shared references that link the required behavior to the chosen provider, its execution resources, and the authority and terms governing their use.

\begin{figure*}[!t]
\centering
\includegraphics[width=7.1in]{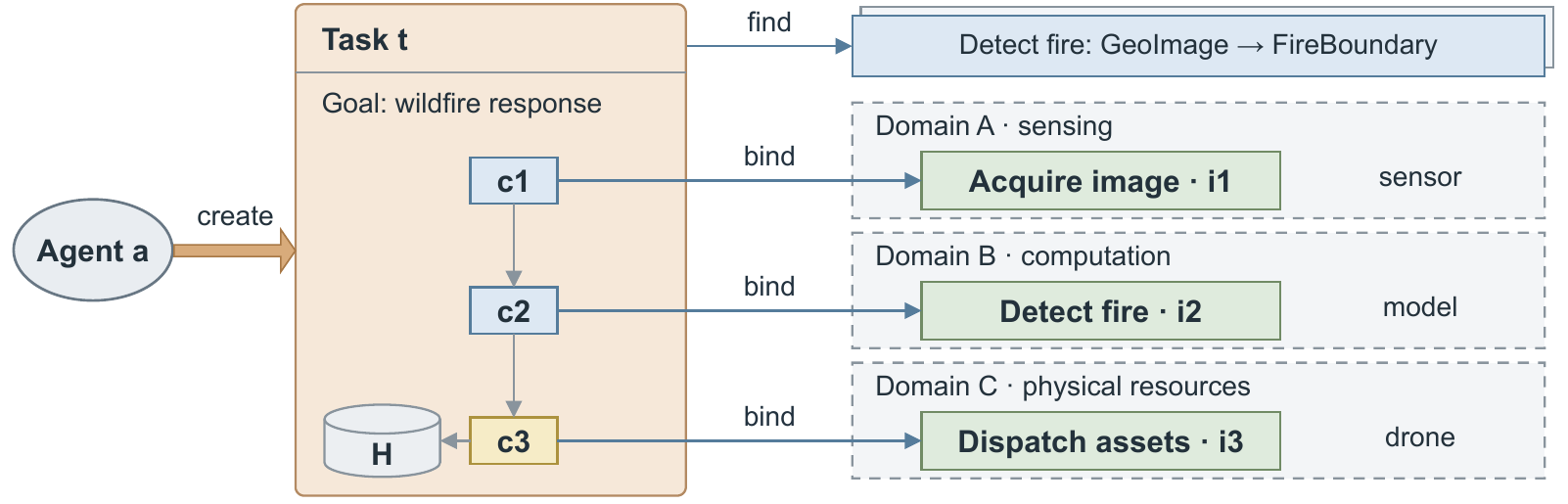}
\caption{Agent, capability, and task relationships in the wildfire response. An agent creates a task whose capability dependencies bind to providers across administrative domains. Stacked records represent capability descriptions, and H denotes task history.}
\label{fig:objects}
\end{figure*}

We propose Capability Internet (CI), an Internet architecture organized around agents, capabilities, and tasks (Fig.~\ref{fig:objects}). Agents request, provide, and coordinate capabilities. A capability is an operation or action offered by a provider under specified conditions. Tasks preserve the goals, dependencies, bindings, state, and history of joint work. Together, these objects connect what a participant needs to what others can supply and to the work already performed.

This model connects computing, intelligence, governance, and value exchange within the same cooperation process. Federated compute resources supply execution capacity, while models, knowledge, tools, and experience supply reusable intelligence. Identity, delegation, allocation, and settlement relate their use to responsible participants and accepted terms. A persistent task carries these relationships through execution, allowing completed work to support later stages and reviewed experience to support future tasks.

We define the Agent--Capability--Task model and five infrastructure functions, then use them to specify how a task selects capabilities, adapts to provider changes, and accounts for their use. We illustrate how CI could use existing protocols through four deployment profiles that inform the evaluation plan.

\section{Architectural Requirements and Related Work}
\label{sec:motivation}

\subsection{Requirements of Agent Cooperation}

Cross-provider tasks need an identity and record that outlast any individual invocation. A production or logistics task can continue through planning, reservations, execution, and delivery, with different agents responsible for different stages. Its goal, dependencies, completed work, and remaining obligations must therefore remain available as participants and connections change. This continuity makes the task a unit for coordinating independently supplied capabilities.

Task continuity also spans administrative boundaries. Providers use different identity systems, operating regions, data policies, and allocation rules. Cooperation must relate each contribution to the requesting principal, the authority for the action, and the conditions governing data and resource use. Shared identity, delegation, and task references let organizations apply their own controls to the same continuing exchange.

Resource discovery must distinguish reusable behavior from current execution conditions. The same inference operation may be available from cloud, edge, or device providers with different capacity, locality, and network conditions. Models, knowledge, tools, and memory can likewise be supplied as versioned resources used by many tasks. Descriptions identify the behavior or resource sought; current availability and policy determine which provider can supply it. Retaining the selected versions also gives later stages and future tasks an interpretable basis for reuse.

Capability use also has an allocation and economic dimension. A task may obtain compute, model access, or physical services through quotas, priorities, subscriptions, or priced offers. Its records must connect the selected terms to authorized use and the resulting allocation or settlement. Persistent tasks, cross-domain authority, resource provision, intelligence reuse, and attributable exchange together define the requirements addressed by CI's infrastructure functions.

\subsection{A Task That Crosses Providers}

\begin{figure*}[!t]
\centering
\includegraphics[width=7.1in]{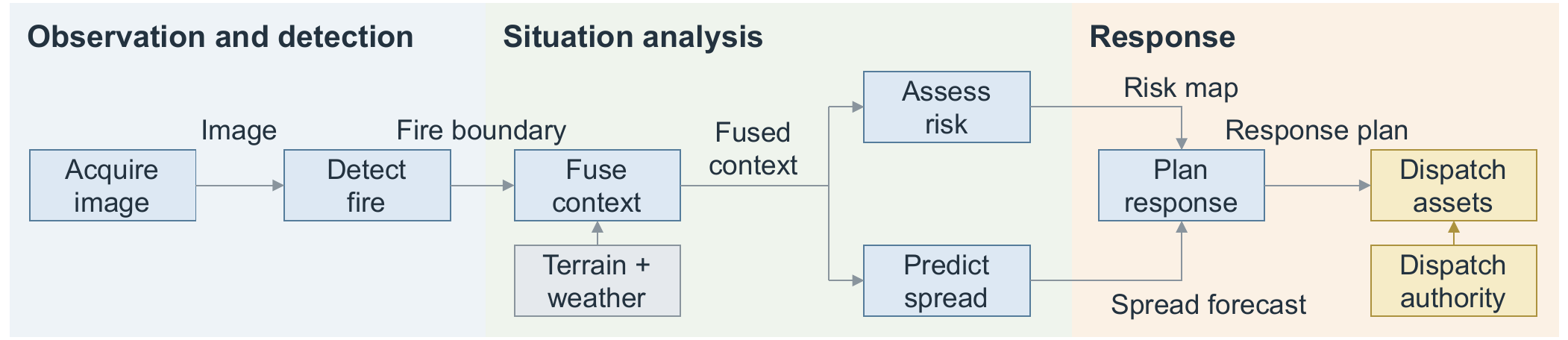}
\caption{Wildfire-response task graph. The acquisition step covers satellite observation and image delivery. Risk assessment and spread prediction share fused context and jointly supply response planning.}
\label{fig:wildfire}
\end{figure*}

A regional coordinator responding to a wildfire obtains a georeferenced image from an imaging satellite. The satellite prepares the image onboard and the scheduled downlink delivers it to the ground. A terrestrial service detects the fire boundary, which is then combined with weather and terrain data. Risk assessment and spread prediction run in parallel on the fused context. Response planning combines their outputs before dispatching authorized assets. Figure~\ref{fig:wildfire} shows these dependencies. Its image-acquisition step includes the orbital observation and delivery path detailed in Section~\ref{subsec:space-ground}.

Each step brings different requirements to the binding decision. Imaging depends on a view of the target, and downlink depends on a ground contact. Detection needs a supported image format and sufficient accelerator memory. Fusion requires recent observations in a consistent geospatial reference, and dispatch requires available assets and authority to act. Section~\ref{sec:execution} follows the execution of these steps under the same task reference.

\subsection{Network, Service, and Agent Foundations}

Network and service research supplies principles for function placement, resource naming, and service composition. End-to-end arguments explain how application requirements guide function placement \cite{ref29}; content-centric networking organizes retrieval around named information \cite{ref2}. Service-oriented architectures establish descriptions, requester--provider interaction, orchestration, and management \cite{ref30}. Semantic Web Services add machine-interpretable conditions, effects, and goal-directed composition, including physical-device interactions \cite{ref31}.

Service steering distinguishes the service requested from the resources selected to deliver it. Service Function Chaining separates an abstract chain from its rendered instance path \cite{ref11}; Intent-Based Networking connects goal expression to translation, activation, and continuing assurance \cite{ibn2023}. Computing-Aware Traffic Steering (CATS) then uses network and compute information for service-instance selection within one service provider \cite{ref13}. CI carries this distinction between required behavior and its realization into tasks spanning independently managed providers.

A2A defines Agent Cards, skills, messages, artifacts, and stateful tasks. MCP exposes tools, resources, and prompts, and its optional Tasks extension supports asynchronous execution \cite{ref8,ref9}. These interfaces provide candidate protocol bindings and descriptor sources for CI's task and resource model.

\subsection{Agent-Oriented Internet Architectures}

Research on an Internet for autonomous agents is developing from agent collaboration toward shared infrastructure. AutoGen organizes applications through programmable multi-agent conversations \cite{autogen2024}; IoA adds third-party agent integration, dynamic teams, and asynchronous task assignment \cite{ioa2025}. AIOS supplies kernel services for scheduling, context, memory, tools, and access control \cite{aios2025}. These systems establish practical mechanisms for coordinating agents and the resources they use.

At network scale, recent research connects agent discovery and orchestration to incentives and regulation \cite{ioaSurvey2026}. Within this broader architecture, Guo et al. develop capability announcement and task-driven discovery through semantic profiling, incremental indexing, and memory-enhanced retrieval \cite{discovery2026}. The Intelligence Delivery Network studies the placement, reuse, caching, and trust of AI services \cite{ref14}, while Agent-OSI connects agent communication to settlement and provenance \cite{ref16}. These contributions address how participants are found, how resources are supplied, and how their use is accounted for.

CI connects discovery and execution through a task record that persists across changes in provider bindings. The coordinator checks a candidate replacement against the required behavior and policy, revalidates affected dependencies, and records the new binding against the same task. The record also links authorization and resource use to the instance that performed each step, providing a basis for settlement and attribution. Table~\ref{tab:positioning} relates these interfaces to mechanisms developed in prior work.

\begin{table*}[!t]
\caption{Related mechanisms and their roles in Capability Internet}
\label{tab:positioning}
\centering\footnotesize
\renewcommand{\arraystretch}{1.13}
\begin{tabularx}{\textwidth}{@{}p{0.23\textwidth}Y Y@{}}
\toprule
\textbf{Approach} & \textbf{Established mechanisms} & \textbf{CI's coordinating interface} \\
\midrule
SOA / Semantic Web Services \cite{ref30,ref31} & Service descriptions, semantic discovery, composition, and policy & Connect descriptions to live resources, authority, and persistent tasks \\
A2A / MCP \cite{ref8,ref9} & Agent interaction, tools and resources, asynchronous task progress & Relate task dependencies, resource versions, and use records across protocol bindings \\
AutoGen / IoA \cite{autogen2024,ioa2025} & Programmable conversations, third-party agents, dynamic teams, and task assignment & Bind typed dependencies to replaceable digital and physical capability instances \\
AIOS \cite{aios2025} & Runtime scheduling, context, memory, tools, and access control & Federate resource offers and task references across independently managed providers \\
IoA architecture / discovery \cite{ioaSurvey2026,discovery2026} & Capability announcement and retrieval, orchestration, incentives, and regulation & Carry behavioral compatibility, resource admission, and use records through the task lifecycle \\
Agent-OSI \cite{ref16} & Agent communication, settlement, linked receipts and provenance & Connect allocation and settlement to capability use and task history \\
\bottomrule
\end{tabularx}
\end{table*}

\section{Agents, Capabilities, and Tasks}
\label{sec:model}

\subsection{Network Participants and Shared Capabilities}

An Agent is an identifiable principal with a degree of autonomy. It may act for a person or organization, or represent a software service or device. An agent can request, provide, or coordinate capabilities, with several agents contributing to one task and each agent participating in several tasks. These roles define who participates in capability cooperation and who is responsible for each contribution.

A Capability is an operation or action exposed by a provider, such as model inference, knowledge retrieval, observation, resource reservation, payment, or device actuation. A service or device can expose its capability through an adapter. The resource description gives agents a common basis for deciding whether the behavior fits a task and under what conditions it can be used. Capability exchange comprises discovery, authorized invocation, composition, and the associated allocation or settlement.

\subsection{Capability Descriptions and Instances}

A capability descriptor specifies reusable behavior:
\begin{equation}
\begin{aligned}
D_c=\langle &\mathrm{cid},\tau_{\rm in},\tau_{\rm out},\mathrm{Pre},\mathrm{Post},\mathrm{Eff},\\
             &\Sigma,Q,\Pi,L,V\rangle .
\end{aligned}
\label{eq:descriptor}
\end{equation}
The identifier $\mathrm{cid}$ names the capability, and $V$ identifies its description version. Input and output types, preconditions, postconditions, and externally observable effects specify the behavior. The state schema $\Sigma$, quality attributes $Q$, policy $\Pi$, and lifecycle $L$ describe its operational requirements. Domain profiles supply concrete schemas, predicates, effect vocabularies, and measurement rules.

A live capability instance identifies the responsible provider, invocation endpoint, current state, quality observations, lease, and evidence references. Multiple providers can realize the same descriptor version, and instance updates track changes in load and availability. This distinction lets a task bind a step to a different provider while preserving the step's behavioral requirements.

\subsection{Persistent Tasks and Composition}

A Task holds the execution context through which agents combine capabilities:
\begin{equation}
T=\langle\mathrm{tid},\mathrm{owner},\mathrm{goal},X,\Pi_T,G,B,\sigma,H\rangle .
\label{eq:task}
\end{equation}
The task identifier and owner establish its identity and responsible principal. The goal defines success, $X$ contains constraints, and $\Pi_T$ carries policy. The graph $G$ records capability dependencies; $B$ binds nodes to instances; $\sigma$ records progress; and $H$ preserves decisions, observations, and effects. Distributed records use the same task reference and explicit plan versions.

A coordinator can delegate parts of the graph to other agents with the relevant goal, constraints, authority, and budget. Their progress and outputs contribute to the enclosing task. The task context persists through planning, execution, waiting, adaptation, and completion, connecting each contribution to the work already performed.

Composition relates the outputs and effects of one capability to the requirements of another. Sequential steps establish downstream preconditions; parallel steps satisfy their dependencies and shared-resource constraints; alternatives fulfill the same required behavior; iteration follows a profile-defined invariant and stopping rule. In Fig.~\ref{fig:wildfire}, risk assessment and spread prediction share context and produce separate inputs to planning. Appendix~\ref{app:semantics} specifies instance records and the compatibility relations used for composition and replacement.

\section{Functional Architecture}
\label{sec:architecture}

\begin{figure*}[!t]
\centering
\includegraphics[width=7.1in]{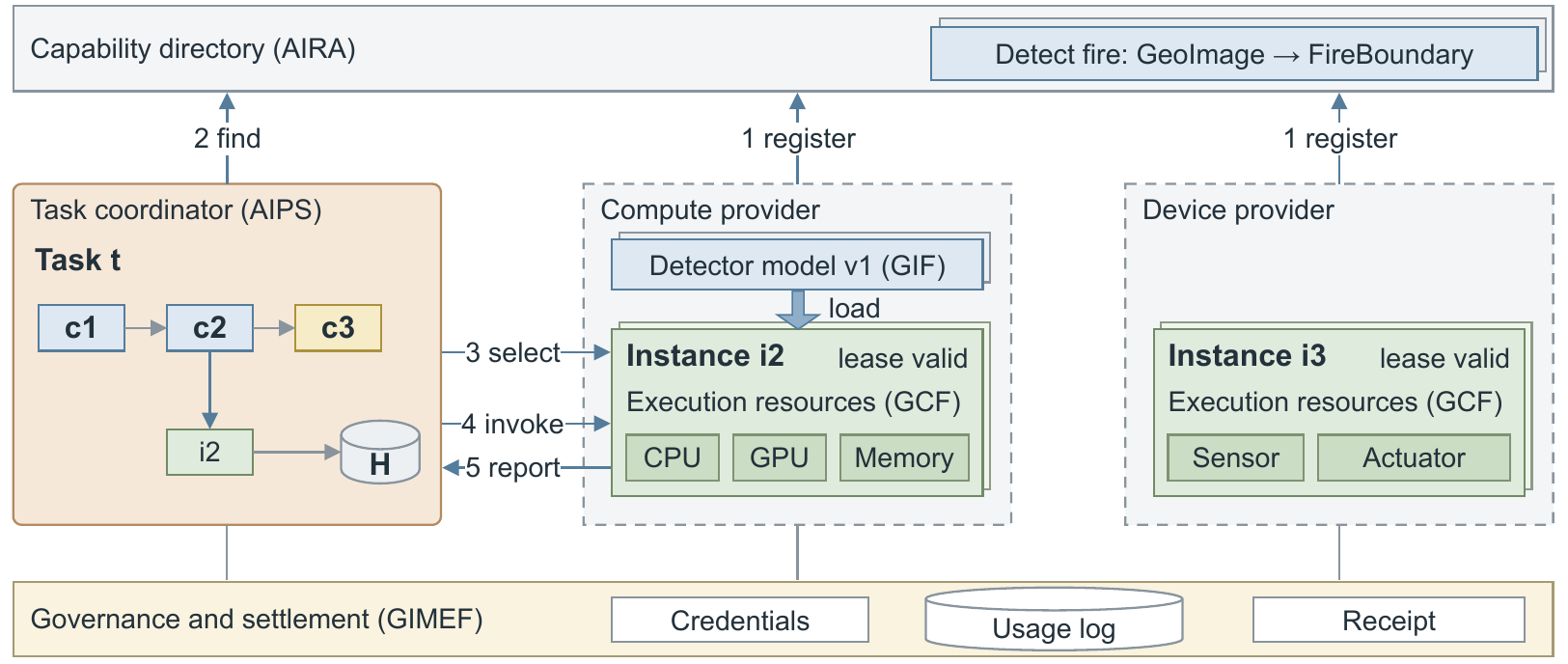}
\caption{Cooperating CI functions at the fire-detection step of the wildfire response. AIRA discovers the capability description, AIPS coordinates execution, GIF exposes the detector model, and GCF supplies execution resources. GIMEF links the interaction to identity, policy, and use records.}
\label{fig:architecture}
\end{figure*}

Five cooperating infrastructure functions manage capability descriptions, instances, and tasks (Fig.~\ref{fig:architecture}). The Agentic Intelligent Resource Architecture (AIRA) represents and discovers capabilities, and the Agentic Intelligent Protocol Stack (AIPS) coordinates task interactions. The Global Compute Fabric (GCF) federates execution resources; the Global Intelligence Fabric (GIF) organizes reusable models, knowledge, tools, experience, and memory. The Governance, Identity, and Machine Economy Fabric (GIMEF) connects participation and resource use to identity, policy, allocation, and settlement. Shared description, instance, and task references connect these functions throughout discovery, execution, and exchange.

\subsection{AIRA: Shared Capability Discovery}

AIRA makes capabilities discoverable across provider boundaries through shared identifiers, descriptions, and domain profiles. Agents publish the behavior they supply and search for behavior required by their tasks. Adapters can map descriptions from A2A, MCP, OpenAPI, device interfaces, and service catalogs into this common representation. Registration authenticates the publisher and validates syntax and publication policy; imported fields retain their provenance.

Federated directories exchange descriptor summaries and namespace delegations, while provider domains control visibility and detailed operational data. Discovery returns matching descriptions, and resolution references lead to current candidate instances. Versioning, revocation, and lifecycle notifications keep these references interpretable as providers change. This organization lets agents discover capabilities beyond their existing integrations and participate in a wider resource network.

\subsection{AIPS: Task Interaction and Coordination}

AIPS organizes cooperation around tasks and their capability dependencies. A coordinator decomposes a goal into required behavior, assigns work to providers or other agents, and relates their progress to the continuing task. Negotiation, invocation, progress, results, adaptation, and completion share task, plan, descriptor, and instance references. A2A and MCP provide candidate protocol bindings; HTTP, RPC, messaging, and delay-tolerant transports carry the resulting interactions \cite{ref8,ref9}.

These references connect local execution to joint progress. A result identifies its producing instance, inputs, resource versions, execution time, and policy evidence, so another participant can interpret and use it. Stable invocation identifiers and an agreed idempotency policy associate retries with existing execution records. The task can therefore continue through changing providers and connections while preserving completed contributions.

\subsection{GCF: Federated Compute and Resource Exchange}

GCF organizes cloud, edge, device, network, and orbital execution resources as a distributed supply of capabilities. Providers expose capacity, availability, location, quality, and allocation terms through instances. Common descriptions let agents seek execution for inference, simulation, analytics, and device services across providers. Edge computing contributes the locality perspective \cite{ref21}; federation makes the resulting resource offers discoverable across resource domains.

An inference task specifies its accelerator and memory needs along with location, latency, or energy requirements. GCF supplies matching offers. The coordinator selects an offer, and the provider confirms admission and issues a lease under its local policy. CATS can guide selection within a provider \cite{ref13}. Provider-local systems handle request scheduling and inference-state reuse, as illustrated by Llumnix \cite{llumnix2024} and Mooncake \cite{mooncake2026}.

Compute exchange couples resource selection to authorization, reservation, metered use, and settlement. A task can obtain capacity through institutional quotas, subscriptions, or priced offers, with GIMEF recording the applicable terms. This makes computing a resource that agents can acquire and combine as their tasks progress.

\subsection{GIF: Intelligence Reuse and Evolution}

GIF organizes models, knowledge, tools, experience, and long-term memory for reuse across tasks. ExpeL shows how experience can supply knowledge for later tasks \cite{expel2024}, and A-Mem dynamically links and updates agent memories \cite{amem2025}. To make such resources usable across providers, GIF maintains their versions, provenance, access interfaces, and update relationships. AIRA publishes the capabilities for retrieving, invoking, or updating them, while GCF supplies execution and storage. Tasks record the selected versions so that their results remain connected to the intelligence resources used.

Execution also contributes to this resource supply. Task results and experience can be reviewed under domain policy and incorporated into new knowledge, memory, tool configurations, or model versions. GIF records the resulting versions and their sources, making accepted contributions available to later discovery and use. Repeated tasks can thus accumulate reusable intelligence and support new capability combinations. AIPS coordinates the updates, and GIMEF carries the access rights, attribution, and exchange terms attached to reuse.

\subsection{GIMEF: Governance and Machine Economy}

GIMEF provides shared participation and exchange functions across capability lifecycles. It identifies the requesting principal and responsible provider, relates authority to the requested effects, and records resource allocation and use. Verifiable credentials supply structured claims \cite{ref24}; token exchange supplies delegation and actor-chain mechanisms \cite{ref25}. Domain policies specify execution regions, data handling, onward delegation, and record retention. Federated principal references and task records let organizations apply their own identity, policy, and accounting systems to the same exchange.

Governance becomes reusable infrastructure for capability cooperation. An agent can reserve resources, acquire a service, and authorize its cost under a delegated task budget. Providers can allocate by priority or quota, charge an internal account, or accept commercial payment. GIMEF links these arrangements to the same capability instances, task decisions, and use records. These functions provide a common basis for a machine economy in which autonomous systems exchange capabilities and the value associated with their use.

\section{Task Execution and Value Exchange}
\label{sec:execution}

\subsection{Discovery, Composition, and Selection}

A task starts with the behavior and conditions needed to reach its goal. AIRA retrieves candidate descriptions through schema lookup, ontology relations, registered translation, or semantic search. The coordinator checks types, preconditions, effects, and policy, then resolves matching descriptions to current instances. Availability, resource capacity, quality, authority, and allocation terms determine which offers the coordinator can select, subject to provider admission.

A feasible binding assigns instances to the graph while satisfying dependencies, shared-resource constraints, permitted effects, and budget. In the wildfire task, the binding pairs an observation opportunity with a downlink that can deliver the image in time for ground analysis. The coordinator compares feasible bindings using task-level quality and resource metrics and selects among nondominated options according to task preferences. Multi-constraint routing supplies models for additive and bottleneck metrics \cite{ref23}. CI also considers behavior, authority, and resource use. The task records the selection rule and supporting observations.

\subsection{Continuing Execution across Providers}

AIPS records the selected bindings and initiates execution. Progress events update $\sigma$, and results and material decisions extend $H$. Each result retains references to its inputs, descriptor, instance, and intelligence resources. These records let later stages interpret completed work and use its outputs.

Changes in availability, quality, policy, or task requirements prompt the coordinator to revalidate the affected bindings. A compatible replacement receives a new binding record under the same task. Completed results can be reused while they meet the current dependencies and constraints. If the selected provider cannot take on the fire spread forecast, the coordinator assigns that step to another provider. The new provider can use the data already fused for the task if they are still current and it has permission to use them.

Before reassigning work already in progress, the coordinator confirms that the earlier attempt can produce no further effects and accounts for any effects it committed. Changes to dependencies, translation steps, or permitted effects create a new plan version. Any compensation or mitigation is recorded as a subsequent action \cite{ref34}.

\subsection{Capability Use and Settlement}

Allocation terms established during selection connect a task's functional choices to resource and payment decisions. The task records the capability, allocation method, price or quota rule, validity, and responsible parties under its delegated budget. Invocation and result records then relate these accepted terms to the behavior obtained and resources consumed. This connection provides the evidence needed to account for capability use.

Providers record metered consumption against the task. The agreed pricing or quota rule determines the charge or allocation, and settlement reconciles the resulting obligations and any payments. The same process covers partial completion, cancellation, and provider changes. Logical and causal relationships order the events recorded by distributed participants \cite{ref35}. These linked records support accounting, receipts, and dispute handling. They also connect reusable results and accepted experience to their sources, rights, and contribution history.

\section{Deployment and Evolution}
\label{sec:deployment}

Four deployment profiles expose different conditions that CI must coordinate. Computing and intelligence services emphasize reusable digital resources; manufacturing links planning to physical production; vehicles and robots introduce mobility and handover; space--ground operations introduce intermittent contact. Each profile supplies domain types, availability conditions, and use policies through the same resource and task interfaces.

\subsection{Compute and Intelligence Resources}

Software agents can combine inference, analytics, data access, tools, knowledge, and memory supplied by Internet providers. The task coordinator selects execution offers through GCF according to demand, locality, latency, and allocation terms. GIF identifies the intelligence-resource versions used. Task records connect the resulting outputs to resource subscriptions, organizational quotas, or purchased services. Reviewed results can then update shared knowledge or experience for subsequent tasks.

\subsection{Autonomous Manufacturing}

A production task combines design, simulation, machining, inspection, and logistics across facilities and suppliers. Descriptions express workpiece state, tolerances, reservations, certification, and process effects. Persistent tasks connect digital designs to physical operations and retain completed stages when remaining work is reassigned. Facilities schedule and control their equipment; task records link their contributions to accepted designs, delivered services, and accounting. Validated inspection results can also become reusable production knowledge.

\subsection{Vehicles and Embodied Robots}

A delivery task can combine vehicle navigation, roadside observations, robotic handling, charging, and fleet services from different operators. Capability descriptions expose location, operating conditions, available actions, and authority. Task context carries mission progress across handovers and changing availability, while local controllers execute real-time actions. Models such as OpenVLA provide reusable robot policies \cite{openvla2025}; CI describes the operations a provider exposes through those policies and relates their use to the mission. Vehicle and robot providers can participate through the same discovery and allocation interfaces as software services.

\subsection{Space--Ground Operations}
\label{subsec:space-ground}

\begin{figure*}[!t]
\centering
\includegraphics[width=7.1in]{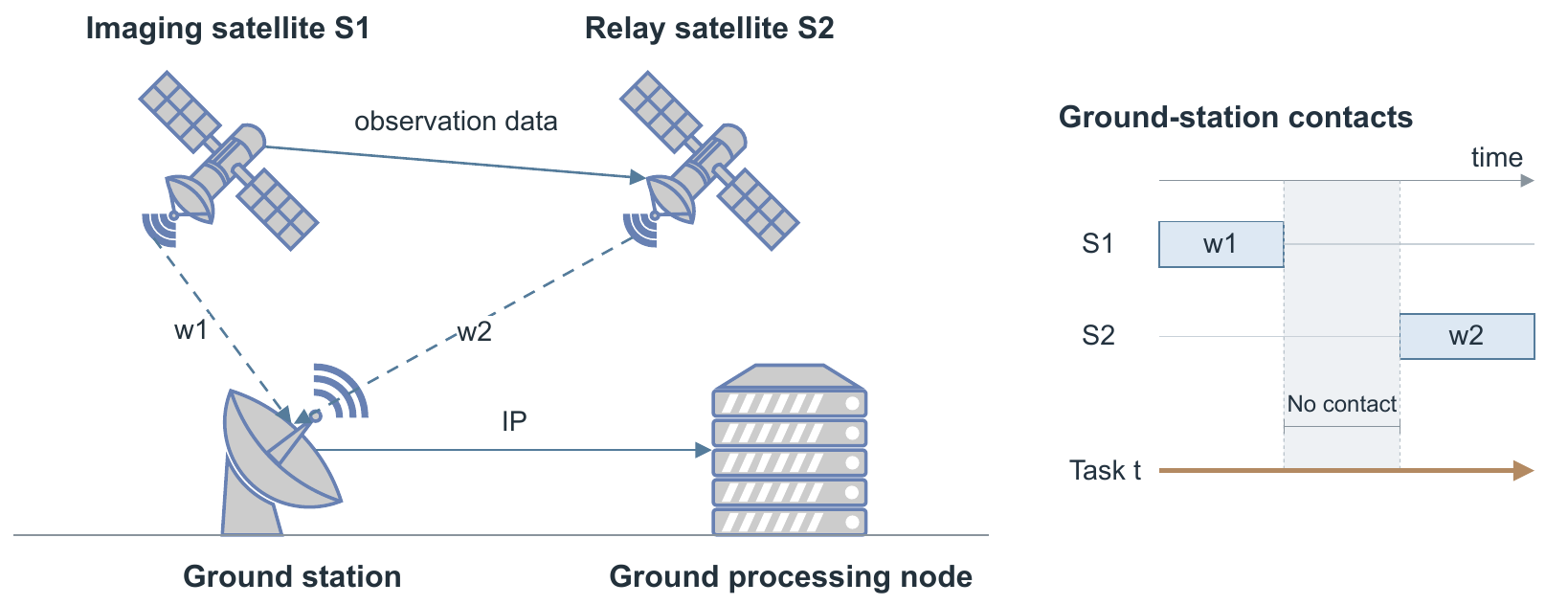}
\caption{The orbital observation and delivery segment of the wildfire task. The imaging satellite supplies the image to the relay, which delivers it through the ground station to a terrestrial processing node. The task persists through the gap between the two ground contacts.}
\label{fig:space}
\end{figure*}

The Space Capability Fabric (SCF) applies CI to observation, navigation, onboard processing, relay, and downlink across orbital and ground providers. Descriptions specify acquisition geometry, contact windows, resources, and mission authority. In Fig.~\ref{fig:space}, the ground station has an earlier contact $w_1$ with the imaging satellite $S_1$ and a later contact $w_2$ with the relay $S_2$. The wildfire task binds acquisition to $S_1$ and delivery to the relay's $w_2$ window. Separate leases let the coordinator plan observation and delivery against their respective availability. The acquisition result stays attached to the task while the delivery binding follows the contact schedule. StarryNet models the satellite dynamics and changing network paths relevant to these choices \cite{starrynet2023}. Bundle Protocol Version~7 provides a store-carry-forward message binding \cite{ref27}.

\begin{table*}[!t]
\caption{Proposed artifacts and measurements for evaluating CI}
\label{tab:evaluation}
\centering\footnotesize
\renewcommand{\arraystretch}{1.13}
\begin{tabularx}{\textwidth}{@{}p{0.27\textwidth}Y Y@{}}
\toprule
\textbf{Evaluation question} & \textbf{Artifact and comparison} & \textbf{Measurements} \\
\midrule
How consistently do profiles represent and compose behavior? & Independent descriptor corpus and compatibility checker; compare conventional service schemas & Matching precision/recall, unresolved fields, false acceptance/rejection, checking time \\
How efficiently does federation resolve current instances? & Multi-domain registry with controlled churn, aggregation, and caching & Query latency, control traffic, candidate recall, measurement age \\
How does task context support continuing execution and placement? & Task engine and resource adapters; compare fixed workflows and provider-local selection & Recovery time, retained work, duplicate effects, deadline satisfaction, decision overhead \\
How are authority and domain policies preserved? & Delegation, data-policy, audit, and failure-injection test suite & Unauthorized-action rate, policy latency, audit completeness, disclosure volume \\
How portable are allocation and settlement? & Two quota, chargeback, or settlement backends using the same task evidence interface & Mapping coverage, reconciliation errors, dispute traceability, integration effort \\
How does execution contribute reusable intelligence? & Versioned intelligence resources and reviewed task outputs; compare resource reuse with separate per-task preparation & Reuse frequency, validation effort, task quality under profile-defined measures, version traceability \\
\bottomrule
\end{tabularx}
\end{table*}

\subsection{Incremental Deployment and Evaluation}
\label{sec:evaluation}

Deployment can start with adapters that map existing catalogs, A2A/MCP interfaces, workflow engines, identity services, and schedulers to descriptions and task records. Federation adds authenticated summaries, lifecycle events, leases, and shared policy and accounting references. Deployment experience can identify message bindings and discovery or resource-steering interfaces suitable for standardization as shared network services.

Table~\ref{tab:evaluation} proposes artifacts and measurements for evaluating the architectural functions. Independently authored descriptions from the four profiles would exercise compatibility, translation, resource contention, effects, and policy decisions. Federated tests would vary catalog scale, instance churn, link conditions, and failure timing. Versioned intelligence resources and reviewed task outputs would provide cases for testing reuse and update traceability. Conformance suites, emulation, and supervised pilots offer successive settings for these evaluations.

Candidate baselines include direct API integration, conventional workflow orchestration, provider-local selection, and native agent-protocol task handling. Recovery comparisons should hold the task graph, candidate providers, and failure timing constant while measuring retained work, recovery time, and duplicate effects. Integration effort and runtime measurements would quantify the cost of coordinating tasks across providers.

\section{Conclusion}

Capability Internet proposes a shared architecture for capability exchange among autonomous agents. Its versioned descriptions connect a task step's behavioral requirements to current provider instances. Persistent tasks link those bindings to authority, progress, and records of use. These links let a task reuse valid results after rebinding and account for each provider's contribution. Reviewed results and experience can also become intelligence resources for later tasks. Four deployment profiles illustrate these interfaces in digital and physical settings and guide their evaluation.

\appendices
\section{Instance and Composition Semantics}
\label{app:semantics}

An instance records the provider and invocation context for descriptor version $V$:
\begin{equation}
\begin{aligned}
I_c=\langle &\mathrm{iid},\mathrm{cid},V,\mathrm{principal},\mathrm{endpoint},\\
            &\mathrm{state},\widehat q,\mathrm{lease},\mathrm{evidence}\rangle .
\end{aligned}
\label{eq:instance}
\end{equation}
The identifier $\mathrm{iid}$ distinguishes implementations, and $\mathrm{principal}$ identifies the responsible provider. The endpoint supplies an invocation reference. The state and timestamped observations $\widehat q$ describe current availability, execution, and quality. The lease bounds availability; evidence references link attestations, telemetry, and provenance.

Under a shared type system, $a\preccurlyeq_T b$ means that a value of type $a$ is acceptable wherever $b$ is required \cite{ref18}. Domain profiles establish this relation through schemas or registered subtypes; a translation capability connects different types through an additional task node. Write $D_a\sqsubseteq D_b$ when the behavior described by $D_a$ can replace that described by $D_b$. Behavioral substitutability \cite{ref19} requires
\begin{equation}
\tau_{{\rm in},b}\preccurlyeq_T\tau_{{\rm in},a},\qquad
\tau_{{\rm out},a}\preccurlyeq_T\tau_{{\rm out},b}.
\label{eq:variance}
\end{equation}
Let $x$ include the initial task-relevant state and input, and $y$ the final state and output. The condition checks are \cite{ref20}
\begin{equation}
\begin{aligned}
\mathrm{Pre}_b(x)&\Rightarrow\mathrm{Pre}_a(x),\\
\mathrm{Pre}_b(x)\land\mathrm{Post}_a(x,y)&\Rightarrow\mathrm{Post}_b(x,y).
\end{aligned}
\label{eq:refinement}
\end{equation}
The replacement delivers the required effects within the set permitted by $D_b$ and the task, and its policy accepts the established authority. Domain profiles define the corresponding quality and lifecycle checks, including units, measurement methods, and validity. These checks relate the replacement's behavior and operating conditions to the task requirements.

For $C_1:\tau_a\!\rightarrow\!\tau_b$ and $C_2:\tau_c\!\rightarrow\!\tau_d$, sequential composition requires $\tau_b\preccurlyeq_T\tau_c$. The first outcome and task context establish the second step's precondition. With $x,y,z$ denoting initial, intermediate, and final task-relevant state and data, successful composition satisfies
\begin{equation}
\begin{aligned}
\mathrm{Post}_{C_2\circ C_1}(x,z)\equiv\exists y\,[&\mathrm{Post}_{C_1}(x,y)\ \land\\
                                              &\mathrm{Post}_{C_2}(y,z)].
\end{aligned}
\label{eq:composition}
\end{equation}
Policy checks govern intermediate-artifact transfer, while effect checks relate the composed execution to task constraints. Quality aggregation follows the structure of that execution: sequential latency can be added, throughput follows its bottleneck rule, and reliability aggregation carries the applicable dependence assumptions. Domain profiles supply the aggregation and measurement rules used to evaluate the composed capability.

\section{Illustrative Descriptor Profile}

The example describes risk assessment. Its domain profile defines types, predicates, measurement methods, signing rules, and version compatibility.

\begin{lstlisting}
capability:
  id: "urn:ci:example:risk-assessment"
  version: "1.0"
  input: "urn:ci:type:fused-fire-context:v1"
  output: "urn:ci:type:risk-map:v1"
  preconditions: ["fresh", "georeferenced"]
  postconditions: ["result conforms to risk-map:v1"]
  effects:
    required: []
    permitted: ["data_release:task-authorized-consumers"]
  quality:
    latency: {unit: ms, aggregation: sum}
    freshness: {unit: s, aggregation: max_age}
  state_schema: "urn:ci:state:invocable:v1"
  policy: "policy_ref"
  lifecycle: "lifecycle_profile_ref"
  resolution: "instance_resolution_ref"
  accounting: "usage_and_settlement_profile_ref"
\end{lstlisting}

\section{Figure Materials}

Figure~\ref{fig:space}: \href{https://github.com/jgraph/drawio/blob/a3cc43efea6e64ecba754f112106b7a7deb5e6d8/src/main/webapp/stencils/networks.xml}{JGraph network symbols} (\href{https://creativecommons.org/licenses/by/4.0/}{CC~BY~4.0}), resized and restyled. Full credit accompanies the sources.

\section*{AI-Assisted Preparation}
OpenAI Codex assisted with language editing, figure preparation, and source checking under author direction. The authors take responsibility for the manuscript.

\IEEEtriggeratref{29}
\bibliographystyle{IEEEtran}
\bibliography{references}
\end{document}